\documentclass{SciPost}

\usepackage[utf8]{inputenc} 
\usepackage[T1]{fontenc} 	
\usepackage[english]{babel} 

\usepackage[bitstream-charter]{mathdesign}

\usepackage{geometry} 		
\usepackage{amsmath} 		
\usepackage{mathtools} 		
\usepackage{float} 			
\usepackage{graphicx} 		
\usepackage{tabularx} 		
\usepackage{booktabs} 		
\usepackage{color, xcolor} 	
\usepackage{pdfpages} 		
\usepackage{extarrows} 		
\usepackage{multirow} 		
\usepackage{multicol} 		
\usepackage{enumitem} 		
\usepackage{xspace} 		
\usepackage{stackrel} 		
\usepackage{tikz} 			
\usepackage{braket} 		
\usepackage{bm} 			
\usepackage{tensor} 		
\usepackage{slashed} 		
\usepackage{siunitx} 		
\usepackage{lastpage} 		
\usepackage{cite} 			
\usepackage[normalem]{ulem} 
\usepackage{fontawesome} 	
\usepackage{tocloft} 		
\usepackage{titlesec} 		
\usepackage{doi} 			
\usepackage{hyperref} 		
\usepackage[most]{tcolorbox} 					
\usepackage[nameinlink, capitalize]{cleveref} 	
\usepackage[nottoc, notlot, notlof]{tocbibind} 	
\usepackage[ruled, vlined]{algorithm2e} 		
\usepackage{makecell}
\usepackage[makeroom]{cancel}
\usepackage{feynmf}
\usepackage{cancel}

\makeatletter
\def\BState{\State\hskip-\ALG@thistlm}
\makeatother

\makeatletter
\@ifundefined{pdfoutput}{}{\DeclareGraphicsRule{*}{mps}{*}{}}
\makeatother

\makeatletter
\DeclareRobustCommand*{\bfseries}{%
   \not@math@alphabet\bfseries\mathbf
   \fontseries\bfdefault\selectfont
   \boldmath
}
\makeatother

\hypersetup{
	pdftitle={ViT4HEP},
	pdfauthor={},
	colorlinks=true, 			
	linkcolor={red!50!black}, 	
	citecolor={blue!50!black}, 	
	urlcolor={blue!80!black} 	
} 

\DeclareSymbolFont{usualmathcal}{OMS}{cmsy}{m}{n}
\DeclareSymbolFontAlphabet{\mathcal}{usualmathcal}

\SetArgSty{textnormal}
\SetKwComment{Comment}{{\small\#}~}{}
\SetCommentSty{mycommfont}

\setitemize{itemsep=0pt, parsep=0pt} 				
\setenumerate{itemsep=0pt, parsep=0pt} 				
\setitemize{itemsep=2pt,topsep=2pt,parsep=0pt,partopsep=0pt,leftmargin=*}
\setenumerate{itemsep=0pt,topsep=2pt,parsep=0pt,partopsep=0pt,labelindent=3pt,leftmargin=*}
\usepackage{amsmath}
 
\usepackage{amsthm} 		
\theoremstyle{definition}

\usetikzlibrary{arrows, shapes.geometric, arrows.meta, shapes, decorations.pathreplacing, fit, patterns, patterns.meta, calc}

\definecolor{Rcolor}{HTML}{E99595}
\definecolor{Gcolor}{HTML}{C5E0B4}
\definecolor{Bcolor}{HTML}{9DC3E6}
\definecolor{Ycolor}{HTML}{FFE699}
\definecolor{Ycolor_light}{HTML}{FFF7DE}
\definecolor{Gcolor_light}{HTML}{F1F8ED}
\definecolor{lightblue}{RGB}{208,232,242}
\definecolor{lightgreen}{RGB}{213,232,212}
\definecolor{lightyellow}{RGB}{255,242,204}
\definecolor{lightlavender}{RGB}{225,213,231}
\definecolor{lightcoral}{RGB}{248,206,204}
\definecolor{lightgrey}{RGB}{230,230,230}
\definecolor{lightpeach}{RGB}{255,230,204}
\definecolor{lightred}{RGB}{255,200,200}

\tikzstyle{expr} = [circle, minimum width=1.8cm, minimum height=1.8cm, text centered, align=center, inner sep=0, draw,font=\LARGE]
\tikzstyle{txt_huge} = [align=center, font=\Huge, scale=2]
\tikzstyle{txt} = [align=center, font=\LARGE, minimum height=1cm]
\tikzstyle{cinn} = [double arrow, double arrow head extend=0cm, double arrow tip angle=130, shape border rotate=90, inner sep=0, align=center, minimum width=2.1cm, minimum height=2.3cm, fill=Bcolor, draw,font=\LARGE]
\tikzstyle{cinn_black} = [cinn, minimum height=2.5cm, fill=black]
\tikzstyle{arrow} = [thick,-{Latex[scale=1.0]}, line width=0.2mm, color=black]
\tikzstyle{loss} = [rectangle, align=center,  minimum width=1.8cm, minimum height=1.5cm,fill=Rcolor,font=\LARGE, rounded corners]
\tikzstyle{xt} = [rectangle, align=center,  minimum width=5cm, minimum height=1.5cm,fill=Gcolor,font=\LARGE, rounded corners]
\tikzstyle{xts} = [rectangle, align=center,  minimum width=1cm, minimum height=1.5cm,fill=Gcolor,font=\Large, rounded corners]

\tikzstyle{embed} = [rectangle, rounded corners=0.3ex, minimum width=1.5cm, minimum height=1cm, text centered, align=center, inner sep=0, fill=Ycolor, font=\large, draw]

\tikzstyle{small_cinn} = [double arrow, double arrow head extend=0cm, double arrow tip angle=130, inner sep=0, align=center, minimum width=1.1cm, minimum height=0.5cm, fill=Bcolor, draw]

\tikzstyle{transformer} = [rectangle, rounded corners, minimum width=6cm, minimum height=2.4cm, font=\large, fill=Gcolor_light, draw]

\tikzstyle{attention} = [rectangle, rounded corners=0.3ex, minimum width=5.5cm, minimum height=1.2cm, align=center, fill=Gcolor, draw, font=\large]

\tikzstyle{crc} = [circle, rounded corners=0.3ex, minimum width=1.5cm, minimum height=1cm, text centered, align=center, inner sep=0, fill=white, font=\LARGE, draw] 

\definecolor{red_cb}{HTML}{e41a1c}
\definecolor{blue_cb}{HTML}{377eb8}
\definecolor{green_cb}{HTML}{4daf4a}
\definecolor{purple_cb}{HTML}{984ea3}
\definecolor{orange_cb}{HTML}{ff7f00}

\definecolor{EmeraldGreen}{HTML}{1ea78d}
\definecolor{EnglishRed}{HTML}{b02427}
\hypersetup{colorlinks=true,urlcolor=EmeraldGreen,citecolor=EmeraldGreen,linkcolor=EnglishRed}

\newcommand{\pl}{p_\text{latent}}
\newcommand{\pd}{p_\text{data}}

\newcommand{\XXLangle}{\biggl\langle}
\newcommand{\XXRangle}{\biggr\rangle}

\def\d{\text{d}}

\newcommand\one{\leavevmode\hbox{\small1\normalsize\kern-.33em1}}
\newcommand{\del}{\partial} 				

\newcommand{\einc}{E_{\text{inc}}} 	

\newcommand{\softmax}{\operatorname{Softmax}}

\newcommand{\geant}{\textsc{Geant4}\xspace}

\newcommand{\arXiv}[2][]{%
	\ifthenelse{\equal{#1}{}}%
	{\href{http://arxiv.org/abs/#2}{arXiv:#2}}%
	{\href{http://arxiv.org/abs/#2}{arXiv:#2~[#1]}}}

\def\slashchar#1{\setbox0=\hbox{$#1$}           
   \dimen0=\wd0                                 
   \setbox1=\hbox{/} \dimen1=\wd1               
   \ifdim\dimen0>\dimen1                        
      \rlap{\hbox to \dimen0{\hfil/\hfil}}      
      #1                                        
   \else                                        
      \rlap{\hbox to \dimen1{\hfil$#1$\hfil}}   
      /                                         
   \fi}

\newcommand{\tikznode}[2]{%
\ifmmode%
\tikz[remember picture,baseline=(#1.base),inner sep=0pt] \node (#1) {$#2$};%
\else
\tikz[remember picture,baseline=(#1.base),inner sep=0pt] \node (#1) {#2};%
\fi}

\def\mathswitchr#1{\relax\ifmmode{\text{#1}}\else$\text{#1}$\xspace\fi}
\def\mathswitch#1{\relax\ifmmode#1\else$#1$\xspace\fi}

\graphicspath{{./figs/}}

\hypersetup{
    colorlinks,
    linkcolor={red!50!black},
    citecolor={blue!50!black},
    urlcolor={blue!80!black}
}

\usepackage[bitstream-charter]{mathdesign}
\DeclareSymbolFont{usualmathcal}{OMS}{cmsy}{m}{n}
\DeclareSymbolFontAlphabet{\mathcal}{usualmathcal}

\fancypagestyle{SPstyle}{
\fancyhf{}
\lhead{\colorbox{scipostdeepblue}{\bf \color{white} ~SciPost Physics Proceedings }}
\rhead{{\bf \color{scipostdeepblue} ~Submission }}

\fancyfoot[C]{\textbf{\thepage}}
}

\begin{document}

\pagestyle{SPstyle}

\vspace*{-2.5em}
\hfill{\small HEPHY-ML-25-05 }
\vspace*{0em}

\begin{center}{\Large \textbf{\color{scipostdeepblue}{
Fast, accurate, and precise detector simulation with vision transformers \\
}}}\end{center}

\begin{center}\textbf{
Luigi Favaro \textsuperscript{1$\star$},
Andrea Giammanco \textsuperscript{1},
Claudius Krause \textsuperscript{2}
}\end{center}

\begin{center}
{\bf 1} Centre for Cosmology, Particle Physics and Phenomenology (CP3), Universit\'e catholique de Louvain, Louvain-la-Neuve, Belgium\\
{\bf 2} Marietta Blau Institute for Particle Physics (MBI Vienna), Austrian Academy of Sciences (ÖAW), Vienna, Austria
\\[\baselineskip]
$\star$ \href{mailto:email1}{\small luigi.favaro@uclouvain.be}\,
\end{center}

\definecolor{palegray}{gray}{0.95}
\begin{center}
\colorbox{palegray}{
  \begin{tabular}{rr}
  \begin{minipage}{0.37\textwidth}
    \includegraphics[width=60mm]{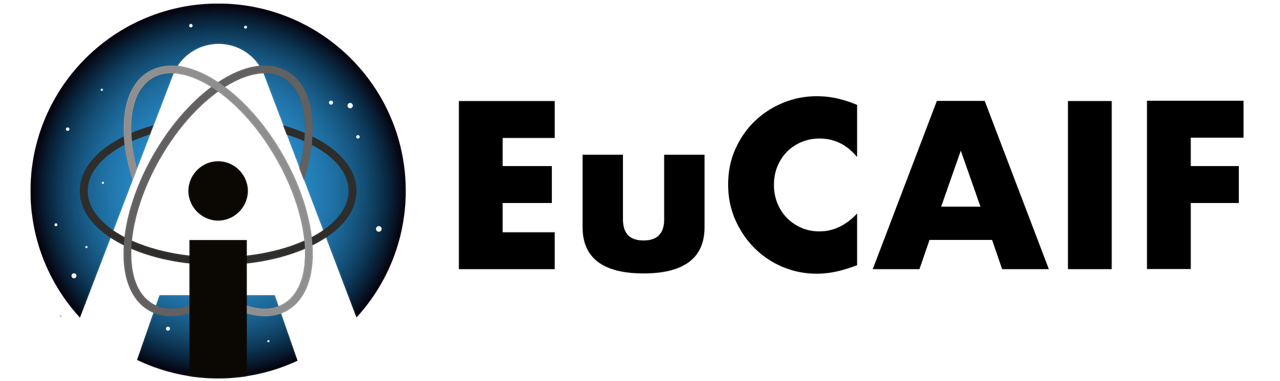}
  \end{minipage}
  &
  \begin{minipage}{0.5\textwidth}
    \vspace{5pt}
    \vspace{0.5\baselineskip} 
    \begin{center} \hspace{5pt}
    {\it The 2nd European AI for Fundamental \\Physics Conference (EuCAIFCon2025)} \\
    {\it Cagliari, Sardinia, 16-20 June 2025
    }
    \vspace{0.5\baselineskip} 
    \vspace{5pt}
    \end{center}
    
  \end{minipage}
\end{tabular}
}
\end{center}

\section*{\color{scipostdeepblue}{Abstract}}
\textbf{\boldmath{%
The speed and fidelity of detector simulations in particle physics pose compelling questions about LHC analysis and future colliders. The sparse high-dimensional data, combined with the required precision, provide a challenging task for modern generative networks. We present a comparison between solutions with different trade-offs, including accurate Conditional Flow Matching and faster coupling-based Normalising Flows. Vision Transformers allows us to emulate the energy deposition from detailed \geant simulations. We evaluate the networks using high-level observables, neural network classifiers, and sampling timings, showing minimum deviations from \geant while achieving faster generation. We use the CaloChallenge benchmark datasets for reproducibility and further development.
}}

\vspace{\baselineskip}

\noindent\textcolor{white!90!black}{%
\fbox{\parbox{0.975\linewidth}{%
\textcolor{white!40!black}{\begin{tabular}{lr}%
  \begin{minipage}{0.6\textwidth}%
    {\small Copyright attribution to authors. \newline
    This work is a submission to SciPost Phys. Proc. \newline
    License information to appear upon publication. \newline
    Publication information to appear upon publication.}
  \end{minipage} & \begin{minipage}{0.4\textwidth}
    {\small Received Date \newline Accepted Date \newline Published Date}%
  \end{minipage}
\end{tabular}}
}}
}




\section{Introduction}
\label{sec:intro}
Precision measurements and new physics discoveries at colliders rely
on a controlled and accurate simulation chain from first principles
to the detector-level observables. However, to match the size of experimental
data collected in the future years, the amount of computing dedicated to
simulations will exceed the available budget. Matching these 
requirements calls for novel, fast, and accurate simulation techniques.
Modern machine learning emulators provide fast, yet precise, alternatives
capable of modeling high-dimensional spaces.\medskip

In particular, generative networks are used to emulate and accelerate 
the calorimetric response in detectors, which often corresponds
to the slowest component of the simulation chain~\cite{Butter:2022rso}. A neural network uses
training data generated with detailed simulations such as \geant~\cite{GEANT4} to approximately learn the distribution
of calorimeter showers and provide an accessible way to sample from
this space. In a large community effort, known as the CaloChallenge~\cite{Krause:2024avx}, many generative networks have been
compared on a set of benchmark datasets. From such a study, 3D Vision Transformers
emerged as a powerful network architecture to model calorimeter showers in
a voxelised space.
We focus on normalising flows, which provide faster sampling but with 
restrictions on the modelling of the data density, and continuous 
normalising flows trained with conditional flow matching, which provide
more flexible transformation at the cost of slower sampling times.
Our discussion on the CFM network is partially taken from CaloDREAM~\cite{Favaro:2024rle},
a submission to the CaloChallenge that showed top-performing results
in several metrics.\medskip

We perform a thorough evaluation of the generative networks in terms of the
sampling time for a calorimeter shower on GPUs, and the fidelity of the generated
showers. Our metrics include layer-wise high-level features and binary neural 
network classifiers for a holistic comparison against \geant on 
high-level features and voxel-level energy depositions.

\section{CaloChallenge datasets}
\label{sec:data}

For our studies, we use four public~\cite{CaloChallenge_ds1_v3,CaloChallenge_ds2,CaloChallenge_ds3}
datasets that have been used for the Fast Calorimeter Simulation Challenge~\cite{Krause:2024avx}. 
They are \geant simulations of a single particle showering in different
calorimeters. The detector has a cylindrical shape with segmentation along
the radial and angular directions.
Each shower provides the incident energy of the incoming particle and the 
energy deposition in each voxel of the detector geometry. \medskip

The two samples contained in Dataset~1 (DS1) simulate central photons and pions.
These simulations were already used for the training of \textsc{AtlFast3}~\cite{ATLAS:2021pzo}.
These two datasets have irregular geometries with layer-wise segmentation in $r\times\alpha$:
\begin{align}
  \text{photons} &\qquad
  8\times1, \; 16\times10, \; 19\times10, \; 5\times1, \; 5\times1 \notag \;, \\
  \text{pions} &\qquad 
  8\times1, \; 10\times10, \; 10\times10, \; 5\times1, \; 15\times10, \; 16\times10, \; 10\times1 \; ,
\label{eq:voxels_ds1}
\end{align}
for a total of 368 and 533 voxels for photons and pions, respectively.
The incident energies are a discrete set of values in $E_\text{inc} = 256~\text{MeV}~...~4.2~\text{TeV}$, in powers of two.\medskip

Datasets~2 and~3 (DS2/3) instead assume a regular geometry with
custom implementation in \geant. The detector has 90 layers of interleaved absorber and active
material. Each pair is grouped into a single layer, totaling 45 voxel-level detector layers. The number of simulated
electron showers is 100,000, produced at $\eta=0$.
The incident energy is log-uniformly sampled in $E_\text{inc} = 1~...~1000$~GeV.
The two datasets contain the same physics; they only differ in the voxelisation:
in DS2 each layer is segmented into $16 \times 9$ angular and radial voxels,
while a DS3 layer contains $50 \times 18$ voxels.
More details on the \geant simulations are available in~\cite{Krause:2024avx}.

\section{Generative Vision Transformers}
The generation process of a single shower is factorised into two
parts, each modelled with a dedicated generative network.
One ``energy'' network generates the total energy deposition in each layer,
while a ``shape'' network generates the normalised voxel energy~\cite{Krause:2021ilc}. 
Expressed in terms of energy-ratio features $u_i$,
the energy network generates $p(u_i|\,\einc)$. The shape
network is additionally conditioned on the truth energy ratios and learns
$p(x|\,\einc,u)$. The final sampling procedure follows
\begin{align}
    u_{i} &\sim p_\phi(u_i|\einc) \notag \\ 
    x &\sim p_\theta(x|\einc, u) \; ,
\end{align}
where $\phi$ and $\theta$ are the learnable weights of the energy network and shape network, respectively. 
We study the trade-offs of two modern generative architectures: discrete Normalising Flows (NFs)~\cite{durkan2019neuralsplineflows} and Conditional Flow Matching networks (CFMs)~\cite{lipman2022flow}.
Then, we describe the transformer-based architecture used for the shape
network: a 3D Vision Transformer (ViT)~\cite{Dosovitskiy2020AnII,Peebles2022ScalableDM}.

\subsubsection*{Generative networks}
A normalising flow builds a map from the data space $\pd$ to a simple 
latent space $\pl$, typically Gaussian. We define the forward map,  
represented by a neural network, as $f_\theta(x)$. The loss function 
minimised in a normalising flow is then the negative log-likelihood defined as
\begin{equation}
\mathcal{L}_\text{INN}= - \XXLangle \log\pl(f_\theta(x)) + \log \left| \frac{\del f_\theta(x)}{\del x} \right| \XXRangle_{x\sim \pd} \; .
\label{eq:b3_cov}
\end{equation}
To directly minimise the logarithm of Eq.~\eqref{eq:b3_cov}, normalising
flows limit the family of possible transformations to invertible $f_\theta$
with a tractable Jacobian.

Given an N-dimensional input vector, coupling blocks reduce the computational cost of calculating the 
Jacobian to $\mathcal{O}(N)$ operations by constructing an upper-triangular matrix. In practice, this is achieved by transforming
half of the input features in each block, followed by a random
permutation of the inputs.
High-expressivity of the neural network is ensured by defining
$f_\theta$ as a binned rational quadratic spline (RQS)~\cite{durkan2019neuralsplineflows}.
An RQS predicts the heights, widths, and derivatives of $m$ bins.
Hence, a neural network predicts a vector $f_\theta(x) \in \mathbb{R}^{(3m-1)\frac{N}{2}}$.
These are processed with a $\softmax$ function to ensure the correct normalisation.
The rational quadratic spline, in each bin $k$, is a function of the input $\xi$ of the form
\begin{equation}
    f_k(\xi) = \beta_{0k} + \frac{\beta_{1k}\xi(1-\xi)+\beta_{2k}\xi^2}
        {\beta_{3k} + \beta_{4k}\xi(1-\xi)} \, .
\end{equation}
The details of the parameterisations, including the calculation of the inverse and
the Jacobians can be found in \cite{durkan2019neuralsplineflows}.
Such normalising flows, or invertible neural networks, allow for 
the sampling of a calorimeter shower in a single inverse pass and,
therefore, they are noticeably fast generators. \medskip

A more expressive alternative is a CFM generative network.
We limit the discussion of CFM networks to the essentials and refer
to the original paper~\cite{lipman2022flow} and CaloDREAM~\cite{Favaro:2024rle} for the details.
A CFM generative network is an example of a continuous normalising flow, which introduces a velocity field $v(x,t)$ to map between representation spaces.
A neural network trains such that $v_\theta(x,t) \approx v(x,t)$, typically by minimising a simple mean squared error loss.
However, as $v(x,t)$ is intractable, the minimisation is recast in terms of the conditional velocity field $v(x,t|x_0)$, with, for instance, a linear trajectory between the two spaces.
While CFM networks are strictly more expressive than discrete normalising 
flows, sampling from a trained network requires solving the 
equation
\begin{equation}
    x(t=1) = x(t=0) + \int_0^1 v_\theta(x, t)\, \d t \;.
\end{equation}
We use standard numerical solvers, such as the Runge-Kutta method, 
which evaluate the network several times, therefore slowing the sampling process.

\subsubsection*{Vision Transformers}
The high dimensionality of voxelised calorimeter data makes it unfeasible
to use a transformer architecture directly. To avoid building an attention
matrix with $N^2$ scaling, we devise a patching scheme that collects
predefined neighbouring voxels in a single patch.
Embedded patches are then passed to several transformer blocks, which
build attention matrices of size equal to the number of patches.
The transformation inside a transformer block can be summarised as
\begin{align}
 x_{\text{h}} &= x + \gamma_{\text{h}} g_{\text{h}} (a_{\text{h}}x + b_{\text{h}}), \notag \\
 x_{\text{l}} &= x_{\text{h}} + \gamma_{\text{l}} g_{\text{l}} (a_{\text{l}}x_{\text{h}} + b_{\text{l}}),
\end{align}
where $g_\text{h}$ and $g_\text{l}$ represent the multi-head self-attention
and the feed-forward transformation, respectively~\cite{Favaro:2024rle}.
The scaling parameters $a$, $b$, and $\gamma$ are predicted during training and depend
on $t$, $\einc$, and the energy ratios $u$.
A final layer projects back to the output of interest. In a NF,
the output is a vector containing the parameters of an RQS for each 
transformed voxel, while in a CFM, we predict a velocity vector for each
voxel of the detector geometry. \medskip

More details on the implementation of 3D vision transformers for the calorimeter
data are available in CaloDREAM~\cite{Favaro:2024rle}.
In the NF setting, half of the patches are used to predict the RQS parameters
of the other half. In each coupling block, we transform both halves of the 
shower before randomly permuting the inputs.
While multiple shuffling options are possible along the spatial dimension,
we limit our study to permutations of entire patches, i.e. along the channel
dimension.
For DS1, we introduce additional voxels to match the binning of the segmented layers and allow for regular patching. While the numbers of additional bins can be reduced to the size of the single patch, we leave it to future work.

\section{Results}
Applications of generative networks for fast simulations rely on the
trade-off between sampling speed and fidelity.
Therefore, we evaluate the generative vision transformers in terms of 
generation time and accuracy compared to \geant.
As the sampling time of the layer energies is a sub-leading term to the total generation time, we employ a more precise CFM generator for the $u$ variables throughout the text.

\subsubsection*{Generation time}
\begin{table}[t]
    \centering
    \begin{tabular}{l c @{\hspace{0.5cm}} c @{\hspace{0.5cm}} c @{\hspace{0.5cm}} c}
    \toprule
     & \makecell{NFs \\ $[\text{ms per shower}]$} & \multicolumn{2}{c}{\makecell{CFMs \\ $[\text{ms per shower}]$}} & \makecell{\geant ($E_\text{inc}$ avg.) \\ $[\text{s per shower}]$} \\ \midrule
     & full gen. & 1 step RK4 & full. gen & ---  \\ \midrule
    DS1-$\gamma$ & 1.93(3) & 2.32(3) & 22.7(3)& --- \\
    DS1-$\pi^+$ & 2.12(5) & 2.76(3) & 28.7(5) & --- \\
    DS2 & 6.9(5) & 2.8(1) & 27.5(5) & \multirow{2}{*}{$\mathcal{O}(10)$} \\
    DS3 & 12.3(5) & 7.7(5) &  125(1) & \\
    \bottomrule
    \end{tabular}
    \caption{Generation time for a single calorimeter shower evaluated on a single A100 GPU with batch size 256. For the NFs, we report the time needed to generate a full shower, while for the CFMs, we indicate timings for both the full generation and a single RK4 step. Using RK4, the generation accuracy converges after 20 steps. The \geant reference is a CPU estimate, averaged over the $E_\text{inc}$ range used for the CaloChallenge DS2/3~\cite{batchingAnna}.}
    \label{tab:gen_time}
\end{table}
The evaluation of the sampling time can depend on several external factors
such as hardware, code implementation, and throughput overheads.
In this study, we report the timing for the forward pass of the neural network
on a single A100 GPU with a batch size of 256. We also include the overhead\
of moving the data from the GPU to the CPU.
Table \ref{tab:gen_time} compares the sampling time of NF and CFM
vision transformers. For NFs, we report the generation time for a single 
shower, while for CFMs, we provide timings for a single step of the RK4 solver.
We observe that the quality of generated showers reaches a plateau
after 20 RK4 steps, which corresponds to 80 function evaluations. As a \geant reference, we quote the average CPU time needed to generate a shower for the CaloChallenge DS2/3~\cite{batchingAnna}.The number of steps needed for CFM networks can be further reduced by
using Bespoke samplers~\cite{Shaul2023BespokeSF, Shaul2024BespokeNS, Favaro:2024rle}, a method which preserves the optimality of initial training
of the generative network.

\subsubsection*{Fidelity}
\begin{figure}[t]
    \centering
    \includegraphics[width=0.325\linewidth, page=16]{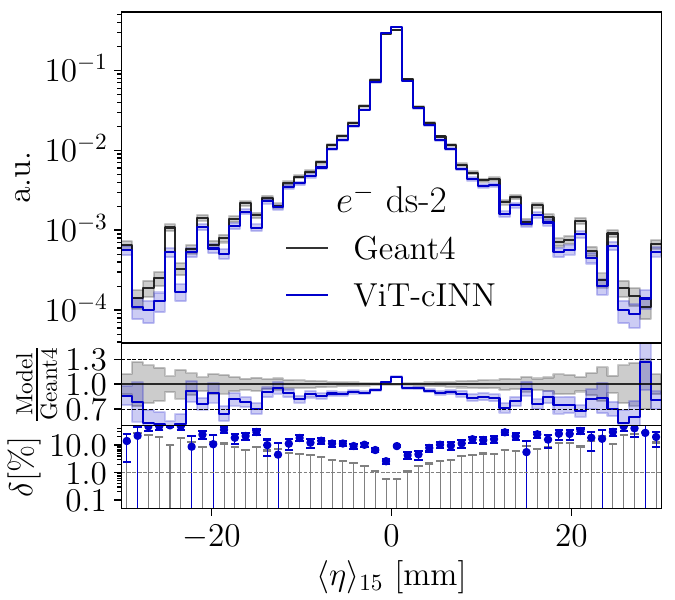}
    \includegraphics[width=0.325\linewidth, page=16]{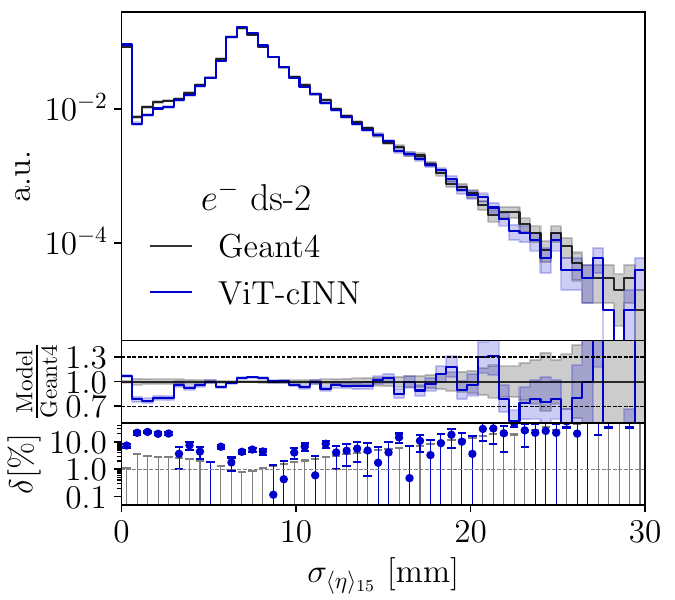}
    \includegraphics[width=0.325\linewidth, page=16]{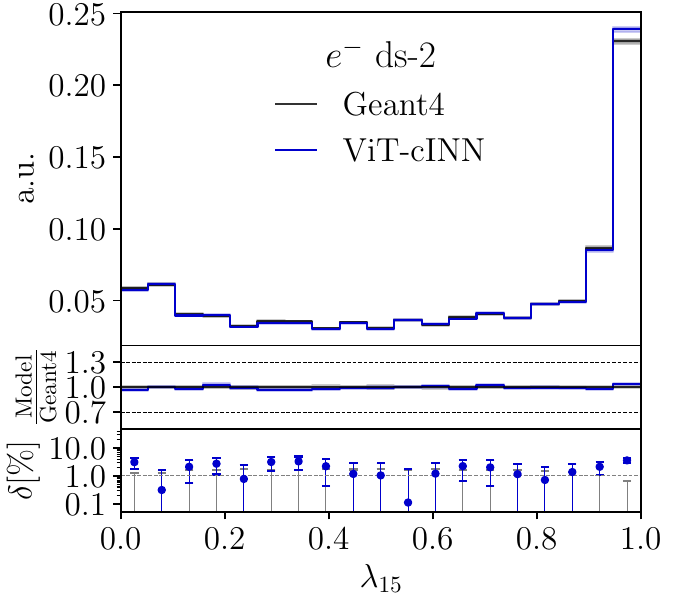} \\
    \includegraphics[width=0.325\linewidth, page=16]{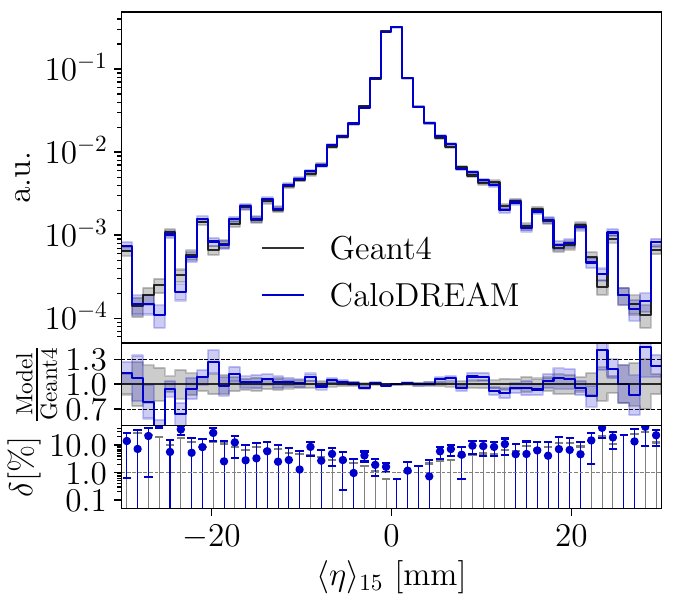}
    \includegraphics[width=0.325\linewidth, page=16]{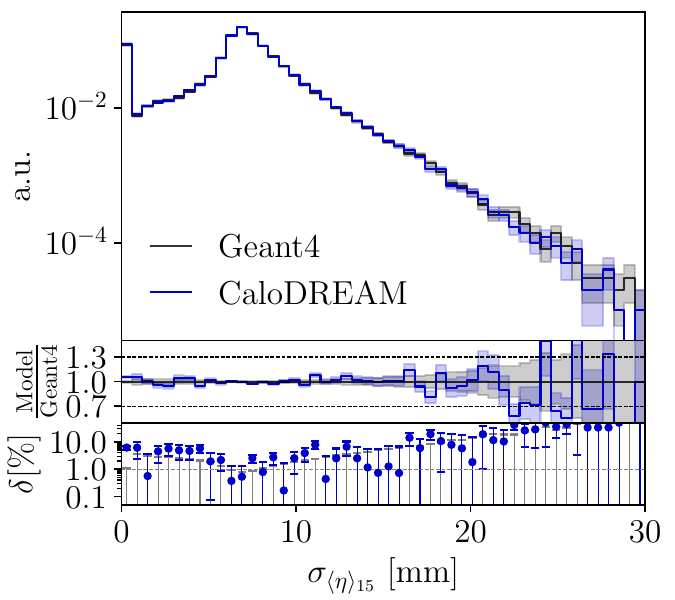}
    \includegraphics[width=0.325\linewidth, page=16]{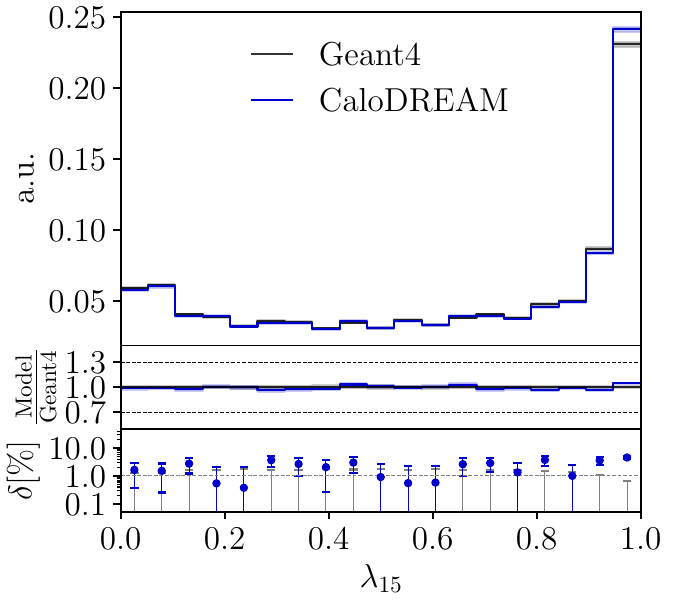} 
    \caption{Set of high-level features for the NF (top) and the CFM (bottom) for DS2. From left to right, we show the center of energy, the width of the center of energy, and the sparsity of calorimeter shower in a single detector layer. CFM plots are reproduced from \cite{Favaro:2024rle}.}
    \label{fig:hlf_ds3}
\end{figure}
A first assessment of the generation quality of the calorimeter shower looks
at high-level features. In Fig.~\ref{fig:hlf_ds3} 
we show the number of inactive voxels, defined as $\lambda$, and two shower 
shape observables. We selected the centre of energy and its width, defined as 
\begin{align}
\langle \eta \rangle 
= \frac{\eta\cdot x}{\sum_i x_i} \; , \qquad\qquad
\sigma_{\langle \eta \rangle} = \sqrt{\frac{\eta^2\cdot x}{\sum_i x_i} - \langle \eta \rangle^2} \; ,
\end{align}
at the layer with the largest average energy deposition as a representative
set of all the high-level features.
A comprehensive evaluation of the generative network should take into
account the high-dimensional correlations of a calorimeter shower.
A neural network classifier trained to distinguish a \geant sample from a generated
one approximates the likelihood ratio of the two distributions,
the optimal test statistic for a simple two-hypothesis test~\cite{Krause:2021ilc,Das:2023ktd}.
Table~\ref{tab:aucs} shows the area under the ROC curve (AUC) score for the networks
trained on the CaloChallenge datasets. An AUC$=0.5$ implies that the generated
samples are indistinguishable from \geant. We report the results of a classifier
trained on the full set of high-level features (HL) and on low-level (LL)
calorimeter data. The classifier architecture and the training 
hyperparameters closely follow those used in \cite{Krause:2024avx}.
We observe that the CFM network has lower AUC scores across the board, while
the NF shows a deteriorating performance as the detector granularity increases.
Additional computer science-inspired metrics that measure the precision and coverage of the generative networks are explored in \cite{Krause:2024avx}.
\begin{table}[b]
    \centering
    \begin{tabular}[t]{l @{\hspace{0.5cm}} c @{\hspace{0.5cm}} c}
    \toprule
     & \multicolumn{2}{c}{AUC} \\ \cmidrule{2-3}
    NFs  & LL & HL \\ \midrule
    DS1-$\gamma$ & 0.61  & 0.53 \\
    DS1-$\pi^+$ & 0.74 & 0.60 \\
    DS2 & 0.68 & 0.78    \\
    DS3 & 0.64 & 0.84  \\
    \bottomrule
    \end{tabular}
    \hspace{1cm}
    \begin{tabular}[t]{l @{\hspace{0.5cm}} c @{\hspace{0.5cm}} c}
    \toprule
    & \multicolumn{2}{c}{AUC} \\ \cmidrule{2-3}
    CFMs  & LL & HL \\ \midrule
    DS1-$\gamma$ & 0.51 & 0.51 \\
    DS1-$\pi^+$ & 0.63 & 0.52 \\
    DS2~\cite{Favaro:2024rle} & 0.53 & 0.52 \\
    DS3~\cite{Favaro:2024rle} & 0.63 & 0.52 \\
    \bottomrule
    \end{tabular}
    \caption{AUC scores from a classifier trained to distinguish \geant from generated calorimeter showers. We train a classifier on high-level features (HL) and low-level energy depositions (LL) for NFs (left) and CFMs (right). The CFM results for DS2/3 are taken from~\cite{Favaro:2024rle}.}
    \label{tab:aucs}
\end{table}
%

\section{Conclusion}
We demonstrated that Vision Transformers are powerful architectures
for fast calorimeter shower generation. For fast simulation applications,
there is a trade-off between sampling speed and accuracy, and the optimal
solution likely depends on the problem at hand.
We presented a fast sampling solution based on RQS normalising flows,
which provides generated calorimeter showers in a single forward pass,
and the more accurate continuous normalising flows trained with CFM,
which can produce samples almost indistinguishable from \geant at the
cost of multiple forward passes needed for a single sample. Our benchmarks, the CaloChallenge datasets, are public and available
to the community for the future development of cutting-edge generative networks.
The code is publicly available at the GitHub repository \href{https://github.com/luigifvr/vit4hep}{https://github.com/luigifvr/vit4hep}. The same repository contains the configurations used to produce these results.

\section*{Acknowledgements}
L.F. is supported by the Fonds de la Recherche Scientifique - FNRS under Grant No. 4.4503.16.
The present research benefited from computational resources made available on Lucia, the Tier-1 supercomputer of the Walloon Region, infrastructure funded by the Walloon Region under the grant agreement n°1910247.







\bibliography{refs.bib}

@article{Krause:2024avx,
    author = "Amram, Oz and others",
    editor = "Krause, Claudius and Faucci Giannelli, Michele and Kasieczka, Gregor and Nachman, Benjamin and Salamani, Dalila and Shih, David and Zaborowska, Anna",
    title = "{CaloChallenge 2022: A Community Challenge for Fast Calorimeter Simulation}",
    eprint = "2410.21611",
    archivePrefix = "arXiv",
    primaryClass = "physics.ins-det",
    reportNumber = "HEPHY-ML-24-05, FERMILAB-PUB-24-0728-CMS, TTK-24-43",
    doi = "10.1088/1361-6633/ae1304",
    month = "10",
    year = "2024"
}

@article{Favaro:2024rle,
    author = "Favaro, Luigi and Ore, Ayodele and Schweitzer, Sofia Palacios and Plehn, Tilman",
    title = "{CaloDREAM -- Detector Response Emulation via Attentive flow Matching}",
    eprint = "2405.09629",
    archivePrefix = "arXiv",
    primaryClass = "hep-ph",
    doi = "10.21468/SciPostPhys.18.3.088",
    journal = "SciPost Phys.",
    volume = "18",
    pages = "088",
    year = "2025"
}

@article{Shaul2023BespokeSF,
  title={Bespoke Solvers for Generative Flow Models},
  author={Neta Shaul and Juan C. P{\'e}rez and Ricky T. Q. Chen and Ali K. Thabet and Albert Pumarola and Yaron Lipman},
  journal={ArXiv},
  year={2023},
  volume={abs/2310.19075},
  url={https://api.semanticscholar.org/CorpusID:264825556}
}

@article{Shaul2024BespokeNS,
  title={Bespoke Non-Stationary Solvers for Fast Sampling of Diffusion and Flow Models},
  author={Neta Shaul and Uriel Singer and Ricky T. Q. Chen and Matt Le and Ali K. Thabet and Albert Pumarola and Yaron Lipman},
  journal={ArXiv},
  year={2024},
  volume={abs/2403.01329},
  url={https://api.semanticscholar.org/CorpusID:268231006}
}

@article{Das:2023ktd,
    author = "Das, Ranit and Favaro, Luigi and Heimel, Theo and Krause, Claudius and Plehn, Tilman and Shih, David",
    title = "{How to understand limitations of generative networks}",
    eprint = "2305.16774",
    archivePrefix = "arXiv",
    primaryClass = "hep-ph",
    doi = "10.21468/SciPostPhys.16.1.031",
    journal = "SciPost Phys.",
    volume = "16",
    number = "1",
    pages = "031",
    year = "2024"
}

@misc{durkan2019neuralsplineflows,
      title={Neural Spline Flows}, 
      author={Conor Durkan and Artur Bekasov and Iain Murray and George Papamakarios},
      year={2019},
      eprint={1906.04032},
      archivePrefix={arXiv},
      primaryClass={stat.ML},
      url={https://arxiv.org/abs/1906.04032}, 
}

@article{Krause:2021ilc,
    author = "Krause, Claudius and Shih, David",
    title = "{Fast and accurate simulations of calorimeter showers with normalizing flows}",
    eprint = "2106.05285",
    archivePrefix = "arXiv",
    primaryClass = "physics.ins-det",
    doi = "10.1103/PhysRevD.107.113003",
    journal = "Phys. Rev. D",
    volume = "107",
    number = "11",
    pages = "113003",
    year = "2023"
}

@article{lipman2022flow,
  title={Flow matching for generative modeling},
  author={Lipman, Yaron and Chen, Ricky TQ and Ben-Hamu, Heli and Nickel, Maximilian and Le, Matt},
  journal={arXiv preprint arXiv:2210.02747},
  year={2022}
}

@misc{CaloChallenge_ds1_v3,
    title = {Fast Calorimeter Simulation Challenge 2022 - Dataset 1 Version 3},
    author = {Michele Faucci Giannelli and Gregor Kasieczka and Claudius Krause and Ben Nachman and Dalila Salamani and David Shih and Anna Zaborowska},
    url={https://doi.org/10.5281/zenodo.8099322},
    howpublished="\url{https://doi.org/10.5281/zenodo.8099322}",
    month={June},
    year={2023}
}

@misc{CaloChallenge_ds2,
    title = {Fast Calorimeter Simulation Challenge 2022 - Dataset 2},
    author = {Michele Faucci Giannelli and Gregor Kasieczka and Claudius Krause and Ben Nachman and Dalila Salamani and David Shih and Anna Zaborowska},
    url={https://doi.org/10.5281/zenodo.6366271},
    howpublished="\url{https://doi.org/10.5281/zenodo.6366271}",
    month={March},
    year={2022}
}

@misc{CaloChallenge_ds3,
    title = {Fast Calorimeter Simulation Challenge 2022 - Dataset 3},
    author = {Michele Faucci Giannelli and Gregor Kasieczka and Claudius Krause and Ben Nachman and Dalila Salamani and David Shih and Anna Zaborowska},
    url={https://doi.org/10.5281/zenodo.6366324},
    howpublished="\url{https://doi.org/10.5281/zenodo.6366324}",
    month={March},
    year={2022}
}

@article{ATLAS:2021pzo,
    author = "Aad, Georges and others",
    collaboration = "ATLAS",
    title = "{AtlFast3: The Next Generation of Fast Simulation in ATLAS}",
    eprint = "2109.02551",
    archivePrefix = "arXiv",
    primaryClass = "hep-ex",
    reportNumber = "CERN-EP-2021-174",
    doi = "10.1007/s41781-021-00079-7",
    journal = "Comput. Softw. Big Sci.",
    volume = "6",
    number = "1",
    pages = "7",
    year = "2022"
}

@article{Peebles2022ScalableDM,
  title={Scalable Diffusion Models with Transformers},
  author={William S. Peebles and Saining Xie},
  journal={2023 IEEE/CVF International Conference on Computer Vision (ICCV)},
  year={2022},
  pages={4172-4182},
  url={https://api.semanticscholar.org/CorpusID:254854389}
}

@article{Dosovitskiy2020AnII,
  title={An Image is Worth 16x16 Words: Transformers for Image Recognition at Scale},
  author={Alexey Dosovitskiy and Lucas Beyer and Alexander Kolesnikov and Dirk Weissenborn and Xiaohua Zhai and Thomas Unterthiner and Mostafa Dehghani and Matthias Minderer and Georg Heigold and Sylvain Gelly and Jakob Uszkoreit and Neil Houlsby},
  journal={ArXiv},
  year={2020},
  volume={abs/2010.11929},
  url={https://api.semanticscholar.org/CorpusID:225039882}
}

@Article{GEANT4,
     author    = "Agostinelli, S. and others",
     title     = "{GEANT4} --- a simulation toolkit",
     journal   = "Nucl. Inst. Meth. A",
     volume    = "506",
     year      = "2003",
     pages     = "250",
     doi       = "10.1016/S0168-9002(03)01368-8",
     SLACcitation  = "%%CITATION = NUIMA,A506,250;%%"
}

@article{Butter:2022rso,
    author = "Badger, Simon and others",
    editor = "Butter, Anja and Plehn, Tilman and Schumann, Steffen",
    title = "{Machine learning and LHC event generation}",
    eprint = "2203.07460",
    archivePrefix = "arXiv",
    primaryClass = "hep-ph",
    reportNumber = "FERMILAB-PUB-22-126-T",
    doi = "10.21468/SciPostPhys.14.4.079",
    journal = "SciPost Phys.",
    volume = "14",
    number = "4",
    pages = "079",
    year = "2023"
}

@misc{batchingAnna,
  author       = {Anna Zaborowska},
  title        = {Level up your performance calculation of the fast shower simulation model},
  howpublished = {\url{https://indico.cern.ch/event/1253794/contributions/5588609/}},
  note         = {ML4Jets 2023},
  year         = {2023}
}


\end{document}